# Giant magneto-optical Kerr enhancement from films on SiC due to the optical properties of the substrate

A. Mukherjee[1], C. T. Ellis[1,2], M. M. Arik[1], P. Taheri[1], E. Oliverio[1], P. Fowler[1], J. G. Tischler[2], Y. Liu[3], E. R. Glaser[2], R. L. Myers-Ward[2], J. L. Tedesco[2], C. R. Eddy Jr[2], D. Kurt Gaskill[2], H. Zeng[1], G. Wang[3] and J. Cerne[1]

[1]Department of Physics, University at Buffalo, SUNY, Buffalo, New York, USA

[2]Electronics Science &Technology Division Code 6800, U.S. Naval Research Laboratory, Washington, DC, USA

[3]Research & Development Center for Functional Crystals, Institute of Physics, Chinese Academy of Sciences, Beijing 100190, China

**Abstract:** We report a giant enhancement of the mid-infrared (MIR) magneto-optical complex Kerr angle (polarization change of reflected light) in a variety of materials grown on SiC. In epitaxially-grown multilayer graphene, the Kerr angle is enhanced by a factor of 68, which is in good agreement with Kerr signal modeling. Strong Kerr enhancement is also observed in Fe films grown on SiC and Al-doped bulk SiC. Our experiments and modelling indicate that the enhancement occurs at the high-energy edge of the SiC reststrahlen band where the real component of the complex refractive index $\tilde{n}$ passes through unity. Furthermore, since the signal is greatly enhanced when $\tilde{n} = 1$, the enhancement is predicted to exist over the entire visible/infrared (IR) spectrum for a free-standing film. We also predict similar giant enhancement in both Faraday (transmission) and Kerr rotation for thin films on a metamaterial substrate with refractive index $\tilde{n} = -1$. This work demonstrates that the substrate used in MOKE measurements must be carefully chosen when investigating magneto-optical materials with weak MOKE signals or when designing MOKE-based optoelectronic devices.

## 1. Introduction

The interaction of light with matter in the presence of an external magnetic field or with a magnetized medium produces a change in the polarization of transmitted and reflected light, known as the magneto-optical Faraday[1] and Kerr [2] effect, respectively. The magneto-optical Faraday and Kerr effects are powerful and sensitive probes of the electronic band structure of materials, revealing fundamental properties such as magnetic anisotropy [3,4], electron spin polarization, [5] and magnetic excitations [6]. Faraday and Kerr spectroscopy in the visible and infrared (IR) spectral ranges have been widely used to study metals [7], magnetic semiconductors [8-11], superconductors [12,13], and more recently graphene [14,15] and topological insulators [16]. These effects also play a vital role in modern technology, for example, Faraday isolators are commonly used to prevent unwanted feedback between devices in telecommunications. The magneto-optical Kerr effect (MOKE) also offers exciting possibilities for ultrahigh density data storage in magnetized media [17], and has recently been used as a powerful tool to probe monolayer ferromagnetism in 2D crystals [18] and magnetoplasmonic effects [19].

In this work, we investigate the enhancement of the MOKE signal at IR photon energies near $E =$ 100 meV, which is achieved in films when the substrate's index of refraction $\tilde{n}$ is near unity. While similar MOKE enhancements have been observed at visible photon energies for magneto-optical films that exhibit $\tilde{n} = 1$ [20], in the present work the enhancement relies upon the optical properties of the substrate rather than the film of interest itself. As such, this greatly improves the flexibility of such an enhancement since it is usually possible to choose a substrate material in order to optimize enhancement effects over a desired spectral range. Furthermore, the prior work on MOKE enhancements in the visible regime rely on the plasma edge induced by free-carriers to obtain $\tilde{n} = 1$. It is well known that these free-carrier based systems exhibit optical losses that are especially strong at IR frequencies. However, in the present case, we exploit the optical properties of polar-dielectric substrate materials, such as SiC, that are able to achieve similar results without the presence of free-carriers. For polar-dielectrics, the $\tilde{n} = 1$ enhancement condition is achieved through the collective oscillations of bound lattice charge that is driven by optical phonons. As such, the optical losses in polar-dielectric materials are much lower for these materials compared to free-carrier systems, which stems from the fact that optic phonon lifetimes are over an order of magnitude larger (~ few picoseconds) than the plasmon lifetimes of metals (~ tens of femtoseconds)[21-23].

Large intrinsic Kerr effects have been observed in various systems such as in transition-metal [24,25] and rare-earth compounds [26]. Mechanisms proposed to explain these large Kerr signals include inter/intraband transitions [27], plasma edge splitting [28], and plasma resonance of charge carriers [20]. Many of these earlier studies address enhancements of the Kerr signal at visible photon energies (1.5-3 eV) [20,27-29]. Some of the previous work [24,25] observed MOKE enhancements down to ~300 meV. Our work focuses on the enhancement of Kerr signal at photon energies near 100 meV. In this present work, we use the optical properties of the substrate to dramatically enhance the MOKE signal from a given material which are often very weak. For example, at visible energies, the saturation Kerr rotation and ellipticity of iron films are of the order of 1 mrad for thickness ranging from 1 to 100 nm [30]. As such, the enhancement of infrared MOKE signals will play a critical role in improving such measurements and the reliability of analyses that rely on such measurements. Furthermore, it was recently suggested that MOKE in graphene films might be exploited to produce ultrafast, IR polarization modulators that are capable of supporting arbitrary waveforms [15]. We expect that the maximum amplitude of polarization modulation that can be achieved in such a system will rely on the exploitation of substrate-mediated MOKE enhancement that is present in the current work.

The origin of the Kerr enhancement can be readily seen in the Fresnel equation for a thick film on a substrate. For a film of thickness $d$ deposited on a substrate with complex refractive index $\tilde{n}$, the complex Kerr angle is given by Ref. [10]

$$\tan\left(\tilde{\theta}_K\right) = \frac{r_{xy}}{r_{xx}} \approx \left(\frac{-2}{Z_0 d}\right)\left(\frac{\tilde{\sigma}_{xy}}{\tilde{\sigma}_{xx}^2}\right)\left[\left(1+\frac{1}{Z_+\tilde{\sigma}_{xx}}\right)\left(1+\frac{1}{Z_-\tilde{\sigma}_{xx}}\right)\right]^{-1}, \qquad (1)$$

where $r_{xy}$ and $r_{xx}$ are the complex reflectivity amplitudes for reflection polarizations perpendicular and parallel, respectively, to the incident light that is polarized in the $x$-direction; $Z_\pm = \frac{Z_0 d}{\tilde{n} \pm 1}$; $Z_0$ is the vacuum impedance; $\tilde{\sigma}_{xx}$ is the complex longitudinal infrared conductivity; and $\sigma_{xy}$ is the transverse (Hall) infrared conductivity. Changes in the real and imaginary components of the Kerr angle correspond to changes in the polarization rotation and ellipticity, respectively. When $\tilde{n}=1$, $Z_- \to \infty$ and dramatic changes occur in $\tilde{\theta}_K$. One way to look at this, is that when $\tilde{n}=1$, the substrate reflectivity goes to zero and the Kerr effect, which is the ratio $r_{xy}$ to $r_{xx}$ (coming from the film and substrate), increases when $r_{xx}$ decreases. For a very thin film on a highly reflective substrate, the Kerr effect can be dramatically reduced as $r_{xx}$ (which is dominated by the substrate) overwhelms $r_{xy}$ (which is primarily coming from the magneto-optical film). It is also interesting to note that for a substrate that has $\tilde{n}=1$ over a wide energy range, the enhancement will occur over that entire range. This means that a free standing film, with vacuum or air on both sides, should exhibit an enhanced Kerr signal at all radiation energies.

## 2. Experimental system and samples

We measure the complex Kerr angle for thin films grown on 4H-SiC substrates. Kerr angle measurements are performed in a magneto-optical cryostat at a sample temperature of 10 K and with magnetic fields up to 7 T that are oriented normal to the sample surface. Linearly polarized, discrete spectral lines produced by $CO_2$ (10.9 – 9.1 μm, 114 – 135 meV) and CO (6 – 5.2 μm, 205-237 meV) gas lasers are reflected by the sample at near-normal incidence. The change in the reflected polarization is measured using photoelastic modulation techniques that employ a ZnSe photoelastic modulator, wire grid polarizer, and liquid nitrogen cooled HgCdTe detector. More details of experimental technique are described in Refs. 8 and 13. Also, we measure the infrared reflectance spectrum of 4H-SiC over the spectral range of $E = 50$-850 meV in order to determine $\tilde{n}$. These measurements are performed at room temperature using a Fourier transform infrared (FTIR) spectrometer that is outfitted with a KBr beamsplitter and HgCdTe detector. A gold mirror is used as a reference to normalize the reflectance spectra.

In this paper we demonstrate MOKE enhancement in three samples: 1) Multilayer graphene epitaxially grown on a c-face 4H-SiC substrate, which was produced by etching the 4H-SiC substrate in hydrogen and then annealing at 1600°C while maintaining a $10^{-4}$ mbar vacuum in a chemical vapor deposition system [31]; 2) Iron films of 3 and 10 nm thicknesses sputtered onto a semi-insulating 4H-SiC substrate, with a sputtering power and pressure of 50 W and 30 mTorr, respectively; and 3) Al-doped SiC prepared in an inductively heated furnace by using a physical vapor transport method [32].

## 3. Results and Discussion

### a) Epitaxial graphene on 4H-SiC

In the presence of an out-of-sample-plane magnetic field $B$, the electronic levels in graphene condense into discrete Landau levels (LLs) [33]. For bilayer and multilayer graphene, the LL energies are proportional to $B$, whereas, for monolayer, the energies are proportional to $\sqrt{B}$ [15]. It is possible to optically excite electronic transitions between LLs when photon energy $E$ matches the energy of an allowed transition between LLs (cyclotron resonance). The selection rules for such transitions allow for two energetically degenerate transitions that are excited by opposite handedness of light of circularly polarized light. Since the MOKE signal is proportional to the difference between the complex, ac conductivity of left ($\tilde{\sigma}_+$) and right ($\tilde{\sigma}_-$) circularly polarized light, these measurements provide a sensitive probe of chiral asymmetries that may exist between these two degenerate transitions. Earlier MOKE measurements on these epitaxially grown multilayer samples have identified over 18 different cyclotron resonances (CR) in a single 0-5 T field sweep [15]. This prior work was able to associate these CR features to distinct monolayers and multilayers with various stacking geometries and attributed the chiral asymmetry to the Pauli blocking of LLs.

Figure 1(a) shows the $B$ dependence of $\mathrm{Re}[\theta_K]$ measured for multilayer graphene grown on a 4H-SiC substrate at $E =$ 120.99, 121.27 and 133.60 meV. By determining the scaling behavior of CR features with $B$ for each $E$, Ref. [15] has shown that the low-field features $|B| < 2$ T follow $\sqrt{B}$ behavior and thus originate from interband LL transitions in monolayer graphene while the resonant energies of CRs features at $|B| > 2\ T$ scale linearly with $B$ and originate from multilayer graphene [15]. As shown in Fig. 1(a), the magnitude of $Re[\theta_k]$ shows a strong dependence on $E$. This is evidenced by not only the magnitude of Kerr features that correspond to CR transitions, but also the linear background component of the signal that results from many high-order overlapping transitions that cannot be distinguished. In order to further explore this dependence on $E$, we focus on the behavior of low field slopes, by performing a linear fit for the data in the range of $|B| \leq 1\text{T}$. While the CR Kerr features may also be used to explore this dependence, doing so is more challenging due to the fact that the lineshape of CR features evolves throughout this range of $E$. At $E = 133.60$ meV (blue curve), the slope is nearly zero (-0.05 mrad/T) and the magnitude of CR features is less than 5 mrad. However, for $E =$ 120.99 meV (black curve), we find an enhancement in the slope (35 mrad/T) and magnitude of the CR features. As shown in Fig. 1(b), the largest enhancement of slope occurs at 120.99 meV (black curve), where the slope of $\text{Re}[\theta_K]$ at its peak is 68 times larger than its corresponding values at higher energies. It is interesting to observe such a dramatic change in slope and magnitude of features for energies differing by less than a few tenths of a meV.

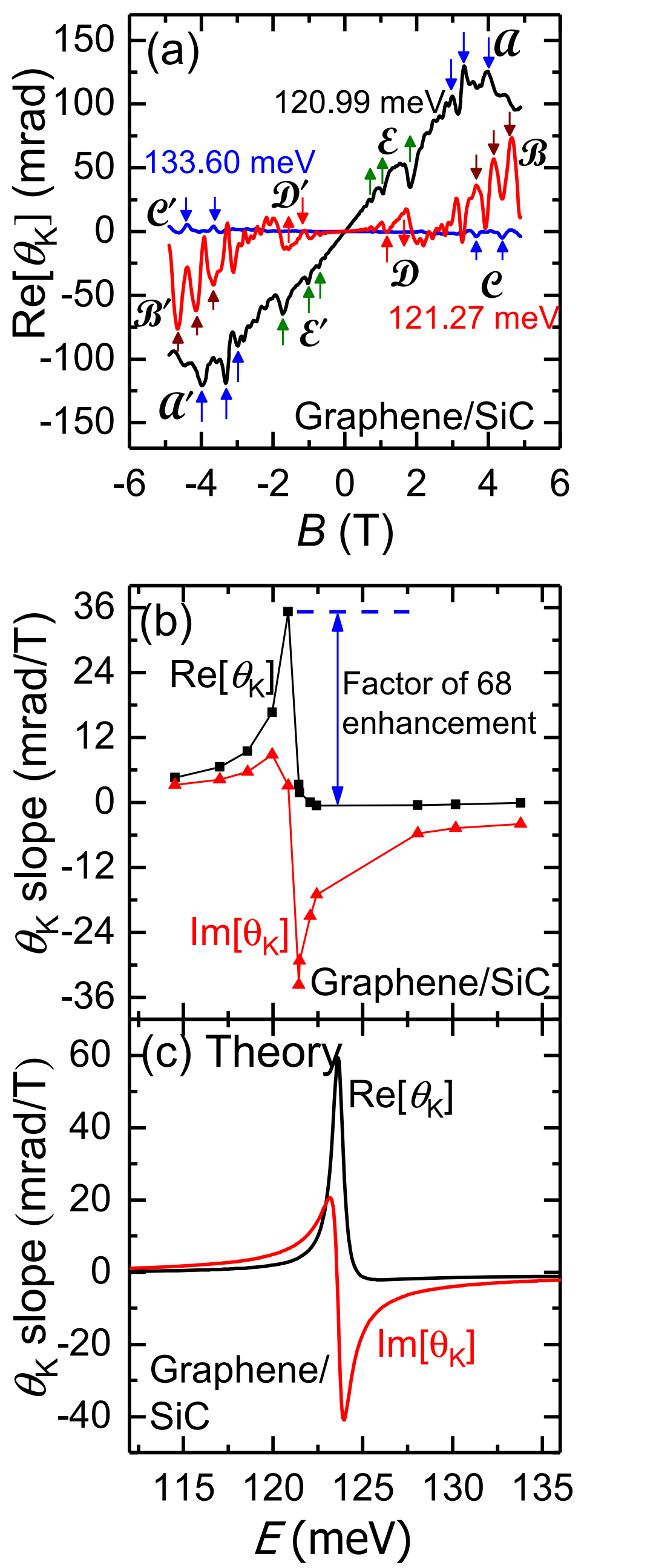

Fig. 1(a) Magnetic field dependence of $\text{Re}[\theta_k]$ for multilayer graphene on SiC at 120.99, 121.27 and 133.60 meV. (b) Energy dependence of low field ($|B| \leq 1$ T) slope. (c) Calculated Kerr ($|B| \leq 1$ T slope) for a monolayer graphene on SiC.

In addition to the $E$ dependent enhancement of the Kerr angle, the lineshape of CR Kerr features also shows an interesting dependence on $E$. In Fig. 1(a) the high-field features are denoted as $\mathcal{A}$ ($\mathcal{A}$'), $\mathcal{B}$ ($\mathcal{B}$') and $\mathcal{C}$ ($\mathcal{C}$') for positive (negative) $B$ at $E$ = 120.99, 121.27, and 133.60 meV, respectively. Arrows indicate the sign of the features, where up arrows represent a dip and down arrows represent a peak in the $Re[\theta_k]$ signal.

The high-field features always exhibit odd symmetry $\theta_K(-B) = -\theta_K(B)$, as indicated by the lineshape inversion that occurs between correlated CR features in the $-B$ and $+B$ regimes (also indicated by the antiparallel orientation of arrows). This is consistent with the expected MOKE signal, which is proportional to the off-diagonal (Hall) conductivity $\sigma_{xy}$ and therefore should have an odd symmetry with respect to $B$ [34]. Interestingly, this is not the case for the observed CR features from monolayer graphene that occur at low-$B$. For $E = 121.27$ meV the typical odd symmetry is maintained for low-$B$ CR features, as indicated by the antiparallel arrow orientations for peaks labeled as $\mathcal{D}$ ($\mathcal{D}$') in the $+B$ ($-B$) regime. However, for $E = 120.99$ meV, an even symmetry $\theta_K(-B) = \theta_K(B)$ is observed for low-$B$ CR features, as indicated by the parallel arrow orientation for dips labeled as $\mathcal{E}$ ($\mathcal{E}$') in the $+B$ ($-B$) regime. This behavior is completely unexpected for MOKE signals. Interestingly, for all spectra measured below and above 120.99 meV, the low-$B$ Kerr signal exhibits CR features with odd and even symmetries, respectively. Furthermore, the crossover from odd to even symmetry features is bounded by the same photon energy (120.99 meV) that results in the maximum Kerr signal. While it is tempting to conclude that both the enhancement and symmetry effects are related to the same mechanism (i.e., $\tilde{n}$ passing through unity, as demonstrated later), these symmetry effects are not captured by our modeling. As such, the origins of such a symmetry change is not understood and remains to be investigated in future studies.

The Kerr angle enhancement observed for graphene on 4H-SiC arises when $\tilde{n}$ passes through unity at the reststrahlen band edge near the longitudinal optical (LO) phonon, which, as shown later, occurs near $E \approx 124$ meV for 4H-SiC. This behavior is captured by eq. 1, where the term $Z_{-} = \frac{Z_0 d}{\tilde{n} - 1}$ becomes large when $n = 1$ and $k \to 0$, resulting in the giant enhancement of Kerr angle values.

The giant enhancement of Kerr effect observed in this work is the long wavelength infrared (LWIR) analog of the MOKE enhancement that is known to occur for metals in the visible spectral range, near the plasma edge [20]. Although there are no free carriers in SiC, and therefore, there is no plasma edge in SiC, throughout the reststrahlen band SiC optically behaves like metal at IR frequencies, as indicated by the high reflectance and the negative-valued real component of the dielectric constant ($\varepsilon_1$). This high reflectance in the reststrahlen band has been the basis behind many novel optical effects in the infrared wavelengths [35-37]. This metal-like behavior is derived from the polar nature of SiC lattice, where there is a charge imbalance between the Si and C atoms. When SiC is illuminated with a frequency between the LO and transverse optical (TO) phonon frequencies that occur in the infrared, the lattice and bound lattice charges are set into motion, effectively screening incident radiation. It is this oscillating bound charge that gives rise to the metallic-like optical nature of the reststrahlen band. Similar to the metals in Ref. [19], $\varepsilon_1 < 0$ for 4H-SiC throughout the reststrahlen band, which gives rise to an effective plasma edge (without free carriers) near the LO phonon energy. However, while the plasma-edge occurs for photon energies in the range of a few eV for metals [20], our phonon driven enhancement occurs at much lower energies. For SiC, this enhancement occurs near 124 meV range due to the LWIR spectral position of the phonons, but by choosing other polar-dielectric materials (e.g., AlN, GaN, and GaAs) this enhancement effect can be pushed to even lower energies that span the terahertz spectral regime. Furthermore, while prior work has shown that Kerr enhancements exist when the magneto-optical material itself satisfies the enhancement condition $\tilde{n} \sim 1$ [20], our work expands upon this, revealing that Kerr enhancement effects also occur in films when the magneto-optical and substrate materials differ and the substrate satisfies the Kerr enhancement condition $\tilde{n} \sim 1$. Further, we would like to emphasize that the large Faraday rotation previously observed in single and multilayer graphene [14] in the terahertz spectral regime arises due to the highly resonant behavior of $\sigma_{xy}$ in graphene at the cyclotron resonance energy. This is in contrast to this work, where the Kerr enhancement is completely independent of $\sigma_{xy}$ or the properties of graphene, and is due to the index of refraction of the SiC substrate going through unity at 120 meV. To obtain a Kerr response, a non-zero $\sigma_{xy}$ is required from the film, which is satisfied by graphene in our system, however, the giant Kerr enhancement is due to the substrate, which is distinctly different than the findings of Ref. [14], which specifically states that the substrate played no role in the Faraday signal. This is further supported by the fact that the work by Crassee et. al. are probing the Faraday rotation at photon energies (10-60 meV) far away from the reststrahlen band of SiC (99-121 meV), where the index of refraction rapidly changes and enhancement is expected to occur.

Figure 1(c) shows the calculated Kerr angle slope for a monolayer graphene on SiC using Eq. 1. The 2D complex conductivity tensor for monolayer graphene was derived by Gusynin et al. using the Kubo formalism [38]. Although the complex conductivities employed in Eq. (1) are 3D conductivities, the 2D formulation of Eq. (1) can be obtained by substituting $\sigma_{ij} = \sigma_{ij2D} / d$, thereafter making Eq. (1) independent of $d$. The sharp enhancement of slope is captured by the model for both $\mathrm{Re}[\theta_K]$ and $\mathrm{Im}[\theta_K]$, as shown in Fig. 1(c), and the overall lineshape of the calculated Kerr angle slope is in excellent agreement with our measurements. For graphene on SiC, the spectral position of the calculated and measured maximum enhancement differ by 2.7 meV, and the measured line shapes are broader than those in obtained via modeling. This may be attributed to the fact that the calculation is only true for a single monolayer. However, in reality, the epitaxially grown graphene sample consists of dozens of layers on the substrate. Also, the calculation assumes graphene to be directly placed on the substrate while the graphene under measurement may be separated from the substrate by many graphene layers that vary in doping level [39]. Further, the model cannot explain the existence of even symmetry features seen for 120.99 meV, as shown in Fig. 1(a) $\mathcal{D}$ ($\mathcal{D}$'). Since the MOKE signal is proportional to $\sigma_{xy}$, which is fundamentally odd in symmetry with respect to the magnetic field, these even symmetry features are unexpected and may be due to contributions from $\sigma_{xx}$, which is even in *B*, to the MOKE signal.

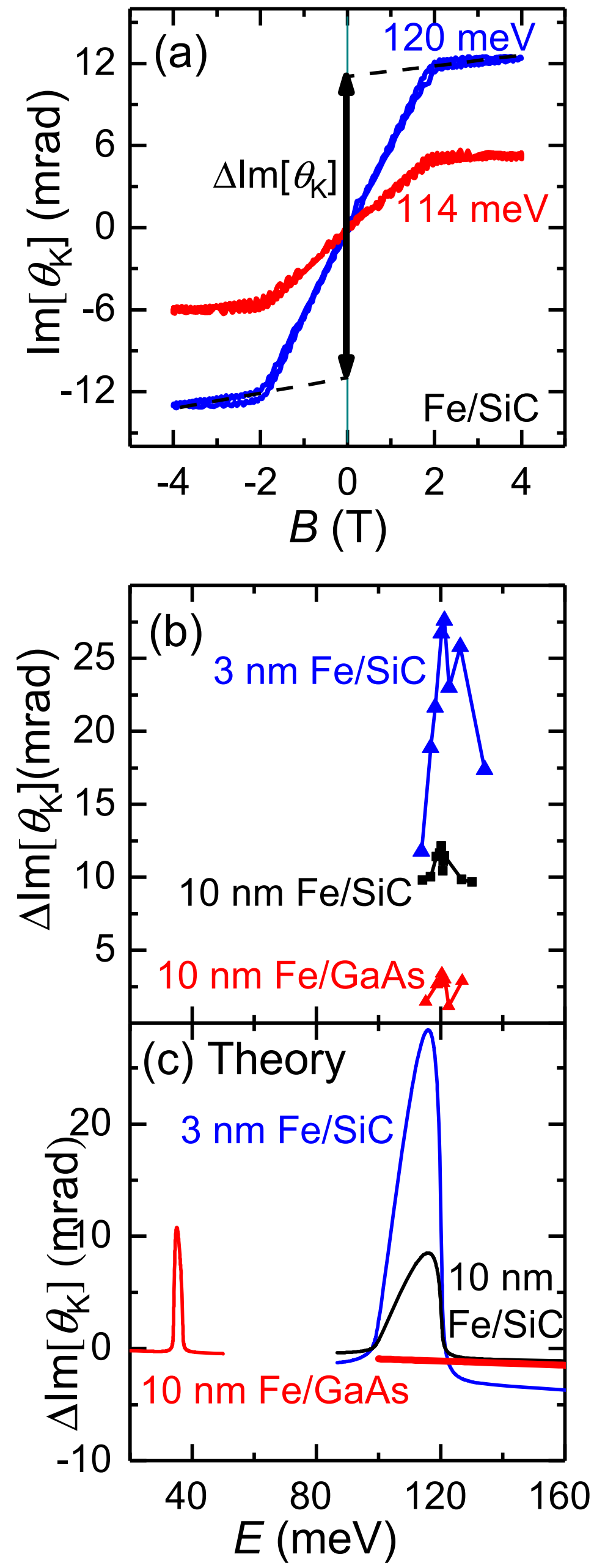

Fig. 2(a) Magnetic field dependence of $\mathrm{Im}[\theta_K]$ for Fe film on SiC, measured at photon energies of 114 and 120 meV. (b) Photon energy dependence of the intercept $\Delta\,\mathrm{Im}[\theta_K]$ measured for various thickness and substrate. (c) Calculated Kerr angle lineshapes for iron films on SiC and GaAs.

### b) Fe film on 4H-SiC

To test these effects more systematically, and determine how they can be translated to other magneto-optical systems, we deposited several iron control films on SiC and GaAs. Iron was chosen since 1) its ferromagnetic behavior enhances the applied magnetic field, producing robust Kerr signals and 2) the hysteretic behavior of the Kerr signal in iron allows the nonlinear contributions from the iron film to be easily separated from linear background signals coming from the substrate and cryostat windows (~ $3\times10^{-2}$ mrad/T). Figure 2(a) shows the results of Kerr measurements performed on a d=3.2 nm thin film of iron sputtered on a 4H-SiC substrate. The ferromagnetic iron exhibits hysteretic and non-linear Kerr angle responses with respect to $B$, saturating near 2 T. The linear hysteresis loop below 2 T is arises from the hard magnetization axis of the Fe film itself, as the magnetization does not saturate until 2 T. There is a small background slope above 2 T, which we have removed by fitting lines to the high *B* field data and tracing them back to *B*=0. For this sample, the Kerr enhancement can be determined by measuring the *E* dependence of $\Delta\,\mathrm{Im}[\theta_K]$ [vertical arrow in Fig. 2(a)], where $\Delta\,\mathrm{Im}[\theta_K]$ is defined as the difference between the *B* = 0 intercepts of fits to the high-field signal where saturation occurs for both +*B* and -*B* [indicated by dashed guidelines in Fig. 2(a)]. An enhancement of $\Delta\,\mathrm{Im}[\theta_K]$ is clearly seen when comparing the response between E = 114 - 120 meV.

Figure 2(b) shows the film thickness and substrate dependence of the Kerr enhancement for thin iron films. For a 10 nm Fe film on GaAs (red curve), $\Delta\,\mathrm{Im}[\theta_K]$ exhibits little energy dependence. However, when the same thickness film is deposited on SiC (black curve), $\Delta\,\mathrm{Im}[\theta_K]$ is a factor of 5 larger throughout the measured energy range. This shows the strong role of the substrate in enhancing the Kerr signal. By reducing the thickness of Fe film on SiC to 3 nm, $\Delta\,\mathrm{Im}[\theta_K]$ is further enhanced by more than a factor of 2, with the peak value near 121 meV.

Equation 1 and a simple Drude model is used to model the MIR Kerr angle response of an iron film on SiC. The ac conductivities of the iron film are given by $\sigma_{xx}(\omega) = \frac{\sigma_0}{1 - i\omega\tau}$ and $\sigma_{xy}(\omega) = \frac{\sigma_0 \omega_c \tau}{(1 - i\omega\tau)^2}$ , where the dc conductivity $\sigma_0 = \frac{Ne^2\tau}{m^*}$, cyclotron frequency $\omega_c = \frac{eB}{m^* c}$, $N$ is the electron density, $e$ is the electronic charge, $\tau$ is the carrier scattering time, $m^*$ is the electron effective mass and $B$ is the magnetic field, which is treated as a fitting parameter. The values from Refs. [40,41] are used for the aforementioned parameters: $N = 17\times10^{22} cm^{-3}$ , $\tau = 1.11\times10^{-15}$, and $m^* = 2.3\times m_e$. For $B = 400$ T, the calculated Kerr angle is in good agreement with our measured values. This value is reasonable for a ferromagnetic sample where the effective magnetic field, as a result of exchange interaction between the electrons are of the order of 100 to 1000 T [42,43].

Figure 2(c) shows the calculated Kerr angle (using eq. 1) for 3 nm and 10 nm Fe films on SiC as well as for a 10 nm Fe film on GaAs. The optical constants for GaAs were taken from Refs. [44,45]. The Fe film on SiC shows enhancement at 120 meV while that on GaAs shows enhancement at 40 meV, which is near the LO phonon of GaAs. With the increase of film thickness, the Kerr angle decreases in magnitude. This behavior is expected since increasing film thickness decreases the amount of light that interacts with the substrate, which is responsible for the enhancement effect.

### c) Al-doped bulk 6H-SiC

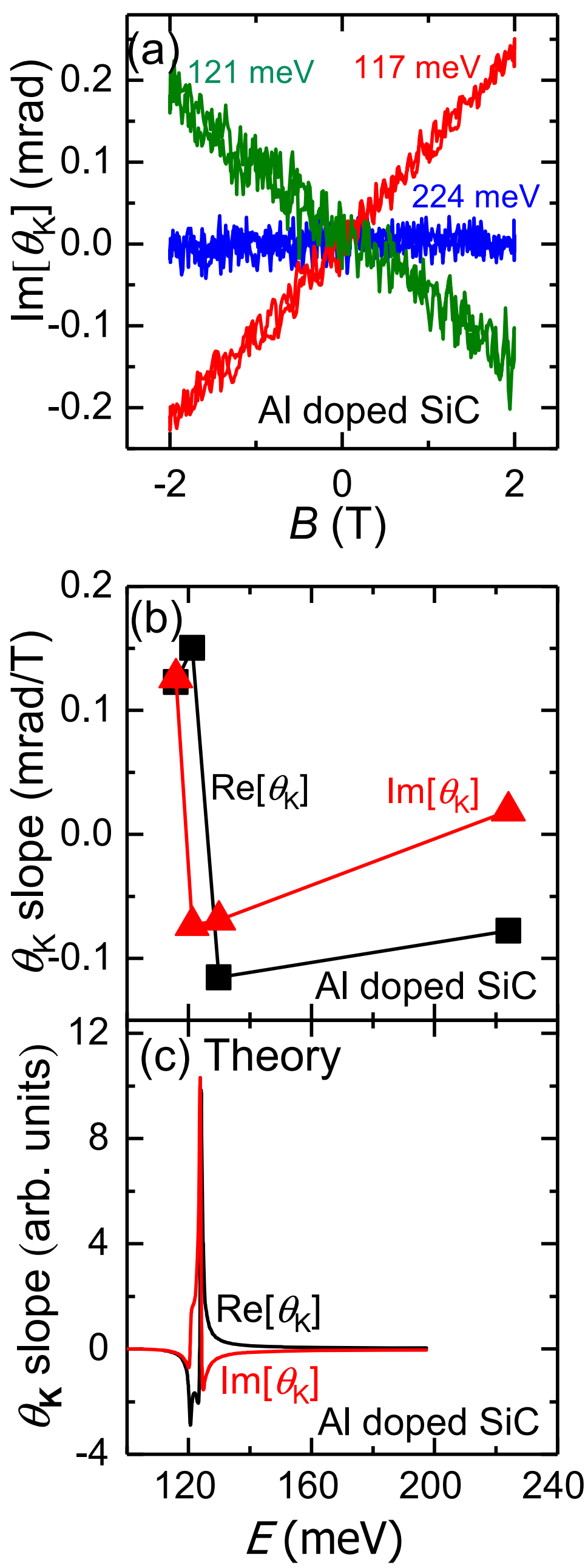

Fig. 3(a). Magnetic field dependence of $\mathrm{Im}[\theta_K]$ for Al-doped 6H-SiC measured at 117, 121 and 224 meV photon energy. (b) Energy dependence of measured slope. (c) Calculated Kerr angle line shape for bulk Al-doped 6H-SiC.

In addition to films on SiC, bulk SiC alloys show similar Kerr enhancement. Figure 3(a) shows the $B$ dependence of $\mathrm{Im}[\theta_K]$ for bulk Al-doped 6H-SiC with $E = 117$, 121 and 224 meV . Although this material shows a ferromagnetic response in its dc magnetization [46], both the infrared $\mathrm{Re}[\theta_K]$ and $\mathrm{Im}[\theta_K]$ are linear in $B$ . This difference can be understood from the fact that magneto-optical response, such as Faraday and Kerr effects, are functions of magnetization, probe energy, and temperature while dc SQUID measurements only probe the magnetization. Similar to the case of graphene, as we tune $E$ the slope of $\mathrm{Im}[\theta_K]$ changes dramatically near 120 meV. At 121 meV (green curve) the slope is negative. When tuned to 117 meV (red curve), the magnitude of the slope increases and also the slope changes its sign, becoming positive. At 224 meV (blue curve) the slope approaches zero and the ellipticity signal is weak.

Figure 3(b) shows the slope of $\theta_K$ as a function of $E$ at 10 K for the Al-doped 6H-SiC sample. A strong energy dependence can be seen for both $\mathrm{Re}[\theta_K]$ and $\mathrm{Im}[\theta_K]$. The sign of slope for both $\mathrm{Re}[\theta_K]$ and $\mathrm{Im}[\theta_K]$ changes very sharply as the photon energy crosses 120 meV. The real and imaginary parts of the Kerr angle for Al-doped 6H-SiC are modeled using the equations for a bulk sample [26] given by

$$\mathrm{Re}[\theta_K] = \frac{4\pi}{\omega}\left[\frac{B\sigma_{1xy} + A\sigma_{2xy}}{A^2 + B^2}\right], \quad (2)$$

$$\mathrm{Im}[\theta_K] = \frac{4\pi}{\omega}\left[\frac{A\sigma_{1xy} - B\sigma_{2xy}}{A^2 + B^2}\right], \quad (3)$$

where $A = n^3 - 3nk^2 - n$ , $B = -k^3 + 3n^2k - k$ and

$$\sigma_{xy} = \sigma_{1xy} + i\,\sigma_{2xy}$$

Since the IR Kerr signal due to the Al-doping is not known, our simple qualitative model assumes a constant value for $\sigma_{xy}$ (i.e. $\sigma_{xy} = 1 + i$) in order to study the effects of the intrinsic SiC. Figure 3(c) shows what one might expect from Al-doped 6H-SiC using the bulk reflectance formulas. Both the model and measurements reveal strong and narrow enhancement features in the complex $\theta_K$ near 122 meV where $n$~1. Unlike the results from graphene and iron films on SiC, the results of our measurements on Al-doped SiC are more akin to those of earlier works [20], where it is the magneto-optical material itself that yields the Kerr enhancement. In both the earlier works and the current work, the enhancement occurs when the magneto-optical, rather than the substrate, exhibits $n \approx 1$. However, in the present case, this effect results from the phonon driven oscillations of bound charge near the SiC LO phonon, rather than the plasma edge that results from the oscillation of free carriers in metals. This shows that even a bulk SiC alloys can produce Kerr enhancement near 124 meV.

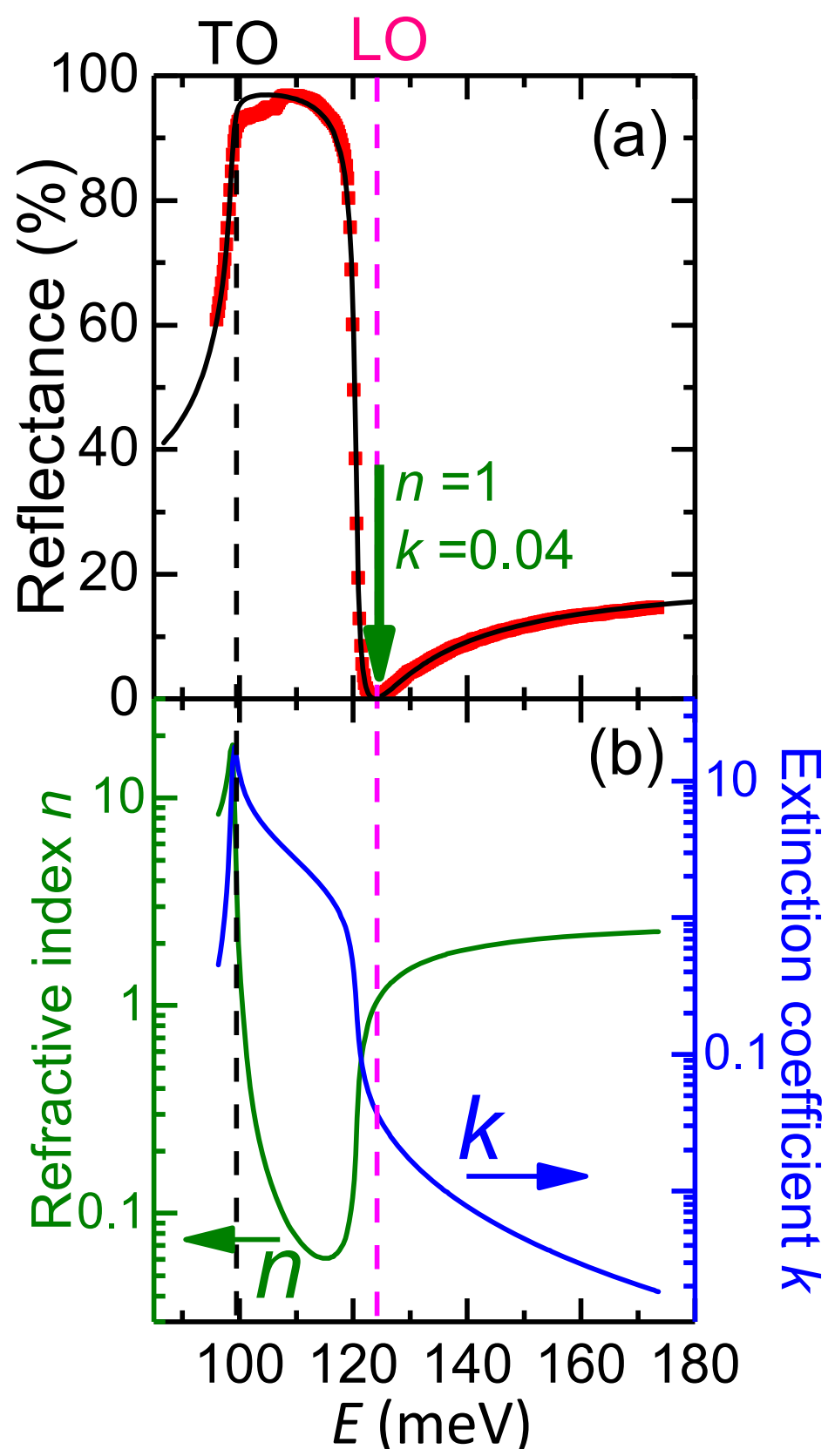

Fig. 4(a) Experimental reflectance (red) of 4H-SiC and calculated fit (solid black line). (b) Refractive index n and k determined from reflectance measurements.

## d) Optical properties of 4H-SiC

Our measuerments and modeling indicate that the optical properties of SiC are critical to the observed Kerr enhancement, so it is important to measure these properties independently to confirm this claim. To determine the optical properties of SiC, the near normal incident reflectance spectrum of 4H-SiC is measured. Subsequently, the complex dielectric function $\tilde{\varepsilon}(\omega)$ and complex refractive index $\tilde{n}$ of 4H-SiC are calculated by modeling the infrared reflectance spectrum (see the Supplemental Material for modeling details). The measured IR reflectance (red symbols) of 4H-SiC is shown in Fig 4(as a function of $E$). The solid curve (black curve) is the calculated fit using the Drude-Lorentz model for dielectric function. The fit-determined energies of the TO and LO phonons are 98.8 meV and 120.5 meV, respectively, which defines the spectral range of the reststrahlen band of 4H-SiC. The photon energy dependence of $n$ and $k$ are shown in Fig. 4(b). To explain Kerr enhancement in our measurements, we only focus on the values of $n$ and $k$ over the measurement range. For 4H-SiC, $n$ passes through unity at two points, first near 100 meV with $k$ = 9.3 and second at 124 meV with $k = 0.0371$. The observed Kerr enhancement occurs when the value of $\tilde{n}$ is unity (i.e., $n = 1$ and $k \to 0$), which according to Eq. 1 is satisfied only at 124 meV.

The Kerr enhancement in 4H-SiC at 124 meV can also be understood from the definition of the Kerr angle, $\tan\theta_K = \frac{r_{xy}}{r_{xx}}$, where $r_{xy}$ is the transverse reflection coefficient and $r_{xx}$ is the diagonal reflection coefficient. An enhancement in $\theta_K$ is expected whenever $r_{xx}$ has a local minimum, assuming that $r_{xy}$ is unaffected by the substrate optical properties and that the magneto-optical film is adequately transparent. Fig. 4(a) shows that at 124 meV, the reflectance of 4H-SiC approaches zero and it is also where we observe the enhancement. This explanation is easier to understand for the case of bulk Al-doped 6H-SiC, where one does not have to account for contributions to the reflectance from both the film and the substrate. For thin films where the reflectance is dominated by the properties of the substrate, this argument should also be valid.

We can extend the analysis to discuss the case of Kerr response from free-standing films by substituting $\tilde{n} \to 1$ (vacuum) in the term $Z_{\pm}$ in Eq. (1). For such a film, the enhancement effect is expected to occur at all frequencies. While our analysis has focused on the reflectance case, our analysis of enhancement effects can also be applied to light that is transmitted through the sample. For example, the Faraday angle for a thin film on a thick substrate is given by [10]

$$\tan(\tilde{\theta}_F) = \left(\frac{\tilde{\sigma}_{xy}}{\tilde{\sigma}_{xx}}\right)\left[1 + \frac{1}{Z_+\tilde{\sigma}_{xx}}\right]^{-1} \tag{4}$$

The term $Z_+ = \frac{Z_0 d}{\tilde{n}+1}$ in the denominator will produce an enhancement of Faraday angle for a metamaterial whose refractive index $\tilde{n}$ passes through $-1$. Such material will also display Kerr enhancement at $\tilde{n} = +1$ due to the presence of $Z_-$ factor in the Kerr angle expression Eq. (1).

## 4. Conclusions

Magneto-optical Faraday and Kerr effects probe the off-diagonal (Hall) conductivity $\sigma_{xy}$, which is very sensitive to the electronic structure of the material. Studying the energy dependence of $\sigma_{xy}$ through Faraday and Kerr measurements provides new information about the energy scale of a system such as plasma frequency, carrier relaxation rate, and cyclotron frequency [47]. In materials showing unusual dc Hall properties, it is especially interesting to explore how the Hall effect behaves at higher frequencies [47-49]. Despite the rich physics that Kerr and Faraday measurements can access, it can be very challenging to measure Kerr signals, which can be very weak and difficult to separate from the background noise. However, our results provide an opportunity to sense these weak Kerr signals by carefully choosing substrate materials that enhance MOKE in the spectral range of interest. We do not claim the largest Kerr signals achieved, but show that the Kerr signal from a given film can be greatly enhanced without changing the properties of the film itself, but by simply placing the film on an appropriate substrate. The reststrahlen band in ionic crystals is known to vary between 7-200 μm ( 200-6 meV) [50,51]. Therefore, even though the enhancement occurs over a narrow energy band, this narrow band can be varied by choice of substrate. Moreover, our model indicates that the enhancement is achievable at all frequencies for free-standing films in air or vacuum. We also predict the existence of giant Faraday rotation for thin films on metamaterial substrate.

## Acknowledgments

Work done at the University at Buffalo, SUNY was supported by NSF-DMR 1410599, by the US Department of Energy, Office of Basic Energy Sciences, Division of Materials Sciences and Engineering under Award No. DE-SC0004890. C.T.E acknowledges support from the National Research Council Postdoctoral Fellowship and Karles Fellowship programs. G. Wang acknowledges support from the National Natural Science Foundation of China (Grant No. 51322211). This work was partially supported by Office of Naval Research.